\documentclass[twocolumn,showpacs,amsmath,amssymb,nofootinbib,superscriptaddress,prl]{revtex4}

\usepackage{graphicx}
\usepackage{dcolumn}
\usepackage{bm}

\begin{document}

\title{Search for Superscreening effect in Superconductor}
\author{P.~Uji\'{c}}
\affiliation{GANIL CEA/DSM-CNRS/IN2P3, Bd H. Becquerel, F-14076 Caen, France}
\affiliation{Vin\v{c}a Institute of Nuclear Sciences, University of Belgrade, Belgrade, Serbia}
\author{F.~de~Oliveira Santos}
\affiliation{GANIL CEA/DSM-CNRS/IN2P3, Bd H. Becquerel, F-14076 Caen, France}
\author{M.~Lewitowicz}
\affiliation{GANIL CEA/DSM-CNRS/IN2P3, Bd H. Becquerel, F-14076 Caen, France}
\author{L.~Achouri}
\affiliation{LPC/ENSICAEN, Bd du Mar\'echal Juin, 14050 Caen Cedex, France}
\author{M.~Assi\'{e}}
\affiliation{Institut de Physique Nucl\'eaire Universit\'e Paris-Sud-11-CNRS/IN2P3, 91406 Orsay, France}
\author{B.~Bastin}
\affiliation{GANIL CEA/DSM-CNRS/IN2P3, Bd H. Becquerel, F-14076 Caen, France}
\author{C.~Borcea}
\affiliation{Horia Hulubei National Institute for Physics and Nuclear Engineering, PO Box MG-6, 76900 Bucharest, Romania}
\author{R.~Borcea}
\affiliation{Horia Hulubei National Institute for Physics and Nuclear Engineering, PO Box MG-6, 76900 Bucharest, Romania}
\author{A.~Buta}
\affiliation{Horia Hulubei National Institute for Physics and Nuclear Engineering, PO Box MG-6, 76900 Bucharest, Romania}
\author{A.~Coc}
\affiliation{CSNSM, Universit\'e Paris-Sud-11 CNRS/IN2P3, 91405 Orsay-Campus, France}
\author{G.~de~France}
\affiliation{GANIL CEA/DSM-CNRS/IN2P3, Bd H. Becquerel, F-14076 Caen, France}
\author{O.~Kamalou}
\affiliation{GANIL CEA/DSM-CNRS/IN2P3, Bd H. Becquerel, F-14076 Caen, France}
\author{J.~Kiener}
\affiliation{CSNSM, Universit\'e Paris-Sud-11 CNRS/IN2P3, 91405 Orsay-Campus, France}
\author{A.~Lepailleur}
\affiliation{GANIL CEA/DSM-CNRS/IN2P3, Bd H. Becquerel, F-14076 Caen, France}
\author{V.~Meot}
\affiliation{CEA/DAM Ile-de-France, BP12, F-91680 Bruy\`{e}res-le-Ch\^{a}tel, France}
\author{A.~Pautrat}
\affiliation{CRISMAT/ENSICAEN et Universit\'{e} de Caen, CNRS UMR 6508, 14050 Caen, France}
\author{M.G.~Saint Laurent}
\affiliation{GANIL CEA/DSM-CNRS/IN2P3, Bd H. Becquerel, F-14076 Caen, France}
\author{O.~Sorlin}
\affiliation{GANIL CEA/DSM-CNRS/IN2P3, Bd H. Becquerel, F-14076 Caen, France}
\author{M.~Stanoiu}
\affiliation{Horia Hulubei National Institute for Physics and Nuclear Engineering, PO Box MG-6, 76900 Bucharest, Romania}
\author{V.~Tatischeff}
\affiliation{CSNSM, Universit\'e Paris-Sud-11 CNRS/IN2P3, 91405 Orsay-Campus, France}

\begin{abstract}
The decay of $^{19}$O($\beta^-$) and $^{19}$Ne($\beta^+$) implanted in niobium in its superconducting and metallic phase was measured using purified radioactive beams produced by the SPIRAL/GANIL facility. Half-lives and branching ratios measured in the two phases are consistent within one-sigma error bar. This measurement casts strong doubts on the predicted strong electron screening in superconductor, the so-called superscreening. The measured difference in screening potential energy is 110(90) eV for $^{19}$Ne and 400(320) eV for $^{19}$O. Precise determinations of the half-lives were obtained for $^{19}$O: 26.476(9)~s and $^{19}$Ne: 17.254(5)~s.

\end{abstract}
\maketitle

The cloud of electrons surrounding the atomic nucleus is known to act as a screening potential. The effect of this screening on a charged projectile is to increase or reduce the penetrability of the projectile through the Coulomb barrier of the positively charged nucleus. Such an electron screening is known to play an important role in astrophysics since it modifies the nuclear cross sections at low energies \cite{assenbaum1987effects}. When measuring the cross sections at energies well below the Coulomb barrier, it is important to take into account the composition of the target since insulator, metallic or alloys materials have different electron screening effects \cite{cruz,raio}. In superconductors, electrons are organized in Cooper pairs obeying the Bose-Einstein statistics. Due to their different nature, Stoppini \cite{Stoppini1991coulomb} predicted that the local density of electrons around the nuclei could be higher than in the normal phase, a strong electron screening  effect, called ``superscreening'', could happen in these materials. It is of high interest to study this predicted superscreening effect because of the possible theoretical implications and because of its potential use in the management of nuclear waste disposal \cite{kettner2006high}.
\par
Electron screening also affects decays of radioactive nuclei \cite{kettner2006high, rose}. From the theoretical point of view, the effects of the environment on the radioactive decay can be reduced to the effect of one parameter called the screening potential energy $U_{e}$. In most cases, it is difficult to calculate the exact value of $U_{e}$ since it depends on several effects, including the electron screening and atomic bonding. Since these environmental effects do not modify the nuclear matrix element, the product of the $\beta$-decay Fermi function $f(Z,Q)$ and the half-life $t$ remains constant. Hence, the change of the $\beta$-decay rate can be calculated as $f(Z,Q')\cdot t'=f(Z,Q)\cdot t$, where $Q'=Q+U_e$ is the modified Q value and $t'$ is the new half-life \cite{rose}. It results in a shorter half-life for the $\beta^{+}$ emission whereas it is longer in the $\beta^{-}$ case \cite{kettner2006high}. When several decay branches are open, the relation $f(Z,Q-E_x)\cdot t^{(i)}=f(Z,Q'-E_x)\cdot t'^{(i)}$ should be used for each individual branch $i$ in order to calculate the new partial half-lives $t'^{(i)}$. It results in a new half-life $1/t'=\sum_i 1/t'^{(i)}$ and new branching ratios $BR'^{(i)}=t'/t'^{(i)}$. Hence, the environment can modify the half-life of the nucleus and also the $\beta$-decay branching ratios (BR). This last effect had not been previously discussed.
\par
In this Letter, we report on the first measurement of the influence of a superconducting host-material on $\beta$-decay. This study was undertaken using intense radioactive beams, high statistics and efficient detectors. In general, high precision half-life measurements are challenging because of the many possible sources of systematic errors, like contamination of the  beam, acquisition dead time, diffusion of the nuclei inside the host-material, pile-up effect and beam or electronic instabilities. The experimental setup was chosen in order to minimize the influence of all these uncertainties on the results.
\par
Two nuclei: $^{19}$O and $^{19}$Ne, were investigated. These nuclei decay by $\beta^-$ and $\beta^+$ with half-lives of 26.464(9) s \cite{audi03} and 17.262(7) s \cite{triambak2012high}, respectively. The isotope $^{19}$O has 3 main decay branches: to the 197 keV level in $^{19}$F with BR of 45.4(15) \%, to 1554 keV with 54.4(12) \% and to 4377 keV with 0.098(3) \% \cite{ensdfonline}. The isotope $^{19}$Ne decays mainly to the ground state of $^{19}$F. An accurate measurement of these properties for nuclei implanted in a superconductor can be used to determine the screening potential energy $U_{e}$ in this material. Conversely, it is possible to predict the effects on half-life and BR if $U_{e}$ is known. In the present experiment, one can suppose that $U_{e}$ is negligible in metal and that $U_{e}$ is equal to the Debye energy in superconductor \cite{Stoppini1991coulomb}. In the case of the Debye plasma model \cite{kettner2006high}, $U_{e}=2.09$x$10^{-11}(Z_t(Z_t+1))^{1/2}(n_{eff}\rho_a/T )^{1/2}$ (eV), with $Z_t$ the charge of the parent nucleus, $n_{eff}\sim$ 1 the number of free-electrons per atom of Niobium, $\rho_a$ the atomic density in units of atoms m$^{-3}$, in Niobium it is $\rho_a = 5.56$x$10^{28}$ atoms m$^{-3}$. Using this formula, one calculates at temperature of 4 K: $U_{e}(^{19}O(\beta^-))=-20.9$ keV and $U_{e}(^{19}Ne(\beta^+))=+25.8$ keV. The Fermi function $f(Z,Q+U_e)$ cannot be calculated analytically, although it could be estimated very precisely by numerical methods \cite{nndclogft}. Using the previous values of $U_{e}$, one obtains the results presented on Table \ref{tab_decay_branching1}. Despite the predicted changes being small, it is measurable and the effect is unambiguous.

\begin{table}
\caption{\label{tab_decay_branching1}Predicted changes of the half-life and BR of ions implanted in a Niobium host in its superconductor phase (4K) relatively to their values in the metallic phase (16K). These values were calculated using differences in screening potential energy of $U_{e}(^{19}O)=-20.9$ keV and $U_{e}(^{19}Ne)=+25.8$ keV.}
\begin{ruledtabular}
\begin{tabular}{lc}
& Expected\\
& change \\
\hline \\
$^{19}$O half-life&+2.5\% \\
$^{19}$O BR to the state at 197 keV&+0.55\%\\
$^{19}$O BR to the state at 1554 keV&-0.24\%\\
$^{19}$O BR to the state at 4377 keV&-14.3\%\\
\hline\\
$^{19}$Ne half-life&-5.0\% \\
\end{tabular}
\end{ruledtabular}
\end{table}

\par
The radioactive nuclei were produced by the SPIRAL facility at GANIL, accelerated to 6 MeV/u and purified using a stripper foil. The purity was checked periodically with a silicon detector. In the case of the $^{19}$Ne beam, the purity was measured to be 99.9(1) \%. In contrast, the $^{19}$O beam had a contamination by 1.23(10) \% of $^{19}$Ne and 45(1) \% of $^{19}$F. The presence of the first contaminant was taken into account in the analysis, whereas the second has no effect on all measurements since it is a stable nucleus. The ions were implanted into a 100 $\mu m$ thick niobium foil. Since the ions are implanted deep into the target, diffusion and leakage of the ions can be neglected. Niobium is a superconducting metal with a critical temperature of 9.2 K. The OptistatCF-V$\mathrm{^{TM}}$ cryostat of Oxford Instruments was used for the cooling of the niobium foil. The temperature stability was monitored for several minutes after each change (see later). No fluctuation larger then 0.5 K was observed.
\par
The detection system consisted of two EXOGAM germanium clover detectors \cite{simpson} and one plastic scintillator detector. The detectors were located between 6 and 10 cm around the target. The gamma-ray detectors were used for the BR measurements. The beta particles were detected with the plastic scintillator and used to determine the half-life. The plastic detector had a thickness of 500 $\mu$m, an area of 5cm$\times$5cm and an absolute efficiency of 3(1) \% to detect the beta particles. A relatively thin detector was used in order to reduce its efficiency to detect gamma rays. The plastic scintillator was connected to two Hamamatsu R2102 photomultipliers. To eliminate the systematic effect called "afterpulsing" \cite{triambak2012high}, the two plastic signals were used in a coincidence mode. The background counting was of $\sim$ 3 counts per second. To limit the effect of the "gain shift" systematic effect \cite{triambak2012high}, the beam intensity was limited to measure a maximum counting rate of 3x10$^{3}$ Hz. In doing so, no significant rate dependence of the half-life was observed. The plastic detector signal was connected to a scaler module through a gate generator with a constant width of 10 ns. Since the scaler module can sustain a frequency of 100 MHz, the dead time of the half-life measurement was practically negligible, at the maximum of the counting rate the correction was of 0.01~\%. The beam implantation time was chosen to last for two half-lives. It was followed by a measurement during which the beam was cut off and the $\beta$-decay measured. The measurement time was chosen to last 20 half-lives in order to search for a temporal variation of the background after implantation. In order to reduce possible systematic effects, the measurement was performed in one-hour cycles. During one hour, several implantation-measurement cycles were performed at low temperature (4K) in the superconducting phase, then the temperature was raised to 16K and several implanttion-measurement cycles were performed in the metallic phase, which was followed by a new one-hour measurement at 4K, and so on. This gave a first set of runs (I) with temperatures 4K and 16K. This experimental approach allowed us to reduce strongly several systematic effects. If, for example, one parameter of the experiment were to change slowly over time, the change would affect equally the two measurements made almost simultaneously and with the same experimental setup. When sufficient statistics was achieved, new cycles were performed with a higher difference in temperature (4K-90K) in order to examine possible temperature dependence of the effect as predicted by the Debye screening model \cite{kettner2006high}. This gave the second set of runs (II) with temperatures 4K and 90K.
\par
The influence of the beam manifests in two ways: a heating of the target which could induce a phase transition from superconducting to normal state, and a lattice damage which changes the critical temperature of the target. The beam power deposited into the target was 10 $\mu$W. In the extreme case, if one supposes the heating is deposited only in the Bragg peak, i.e. inside a target layer of 2 $\mu$m, the specific heat capacity of niobium is 24.6 J mol$^{-1}$ K$^{-1}$, one calculates that the maximal increase of the local temperature is $\approx 1.9K$. This is not sufficient to induce a transition between the superconducting phase at 4K to the metallic phase at 9.2K. The beam irradiation induces a deterioration of the target lattice which results into a change of the critical temperature of the material. Each incident ion induces thousands of vacations and interstitials. Neum\"{u}ller et al. \cite{Neumuller1977446} examined the dependence of the critical temperature of niobium on the irradiation induced by an oxygen beam of 25 MeV, whereby the beam was not stopped inside the target. They found that the critical temperature was decreased by 1\% for irradiation of 1,3$\times10^{16}$ ions of oxygen per cm$^2$. In the present experiment, the beam intensity was around 10$^5$ pps and the total effective irradiation did not exceed 10$^{12}$ ions per cm$^2$ per target. This is much lower than in ref. \cite{Neumuller1977446}, and thus change of the critical temperature can be neglected.

\begin{figure}[t!]
\includegraphics[width=8cm]{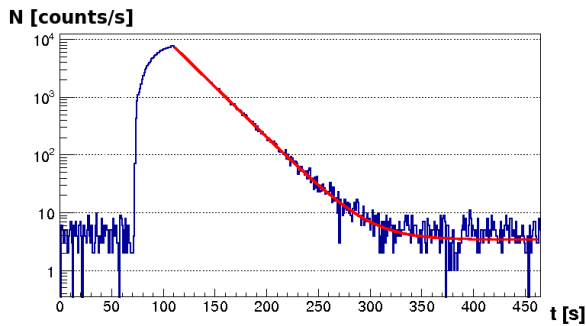}
\caption{\label{fig1}
An example of implantation-decay curve of $^{19}$Ne measured at 4K. The fit of the decay curve was performed using an exponential function plus a constant background. In this example, t$_{1/2}$=17.22 $\pm$ 0.04 s and $\chi^{2}/\nu$ =1.01.}
\end{figure}

\par
\begin{figure}[t!]
\includegraphics[width=8.5cm]{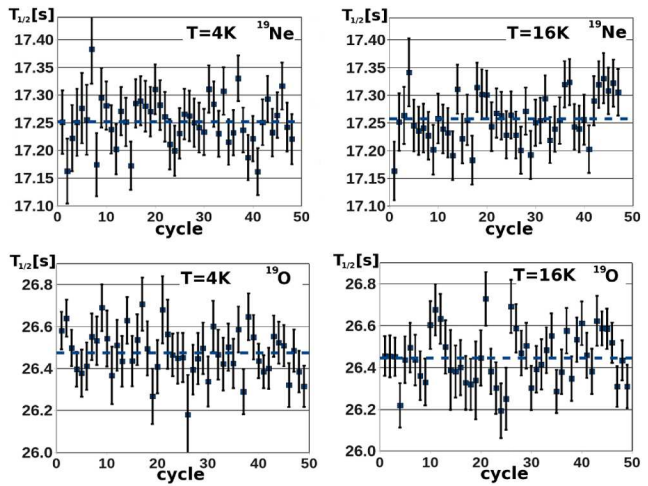}
\caption{\label{fig2}
Upper two figures: measured $^{19}$Ne half-life values for all cycles of measurements of the set of runs (I) (4K -16K). Lower two figures: the same for $^{19}$O. The dashed lines correspond to the mean value.}
\end{figure}
\begin{figure}[t!]
\includegraphics[width=6cm]{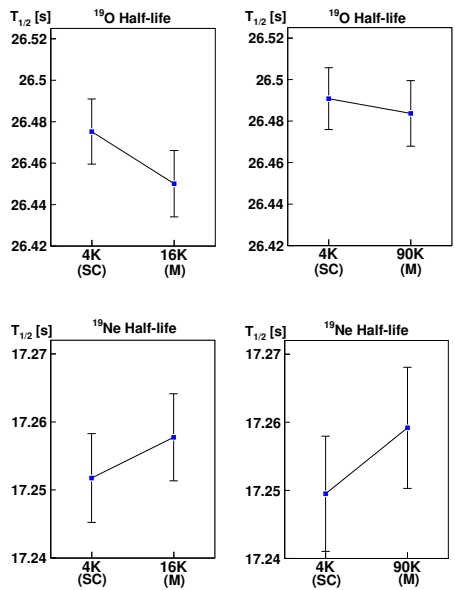}
\caption{\label{fig1:HLBR}
Measured half-lives of $^{19}$O and $^{19}$Ne implanted in niobium at different temperatures.}
\end{figure}

An example of implantation-decay curve is presented in Fig. \ref{fig1}. After each implantation, the decay curve was fitted with an exponential function of time plus a constant. For the fitting procedure, the standard Levenberg-Marquardt method \cite{press2001numerical} was used with additional modifications \cite{baker1984clarification,laurence2010efficient}. The fitting was performed on the full decay curve. On the contrary, only a part of the decay curve was used for the measurement of BR with the Germanium detectors. In this case, the analysis started each time on the same counting rate in order to respect the same experimental conditions, especially regarding the pile-up probability. A half-life value was extracted after each implantation and the mean value of the all measured half-lives was obtained by the weighted mean method, see Fig.  \ref{fig2}. This procedure was performed for each temperature and for each beam. The systematic effect due to the choice of the bin width was taken into account in the calculation of the error.
Unambiguously, 11 cycles out of a total of 331 were rejected because they had $\chi^2/\nu > 1.3$. The problem was identified, it was noticed that these runs correspond to ``time-out'' occurrences when the acquisition was stalled for short period of time because of network congestion, with a loss of time consistency. For the other cycles, a $\chi^2/\nu = 1.04 \pm 0.10$ was obtained for $^{19}$O and $\chi^2/\nu = 1.01 \pm 0.10$ for $^{19}$Ne. These values are indicative of statistical consistency among the runs since the theoretically expected variation of $\chi^2/\nu$ is 0.07 for $^{19}$O and 0.08 for $^{19}$Ne. The results for half-lives are presented in Fig. \ref{fig1:HLBR} and the results of the relative branching ratios are given in Table \ref{tab_decay_branching_exp1}.
\begin{table}
\caption{Measured values of the relative branching ratios (BR) of $^{19}$O in superconducting (SC) and metallic (M) phases of niobium for two sets of runs: (I) 4K and 16K,  and (II) 4K and 90K.}\label{tab_decay_branching_exp1}
\begin{ruledtabular}
\begin{tabular}{lcc}
Phase  & BR of    & BR of  \\
   &197/1554  & 197/4377 \\
\hline
SC-4K    (I)  & $3.611 \pm 0.015$ &$ 3000 \pm 200$ \\
M-16K (I)  & $3.607 \pm 0.015$ &$ 2700 \pm 170$ \\
\hline
SC-4K    (II)  & $3.595 \pm 0.017$ &$ 3100 \pm 200$ \\
M-90K (II) & $3.582 \pm 0.017$ &$ 2800 \pm 200$ \\
\end{tabular}
\end{ruledtabular}
\end{table}
All measured half-lives and BRs are consistent within one-sigma error bar. Many checks of systematic errors were performed, including dead time effect, contamination of the beam, variation of the $\chi^2/\nu$ values, average value and weighted average value consistency, change of the normalized residuals of the fits, change of the background. No systematic effect could be observed except the one produced by the binning of the spectra, which was of 0.011 \%. The mean value of the measured half-lives is 26.476(9) s for $^{19}$O and 17.254(5) s for $^{19}$Ne. These results are in excellent agreement with previously measured values of 26.464(9) s \cite{audi03} and 17.262(7) s \cite{triambak2012high} respectively.
\par
It is possible to combine all these measured results in order to obtain a more sensible and statistically more accurate result. In order to do so, these results have to be normalized. Table \ref{tab_decay_branching1} was used for this purpose. For example, the measured BR 197/1554 was 3.607 at 16K and 3.611 at 4K. Thus, the measured relative change is +0.11\%. According to Table \ref{tab_decay_branching1}, the predicted change is +0.79\%. This means that 0.11/0.79=+13.9\% of the predicted change was measured. All the measured BRs and half-lives, both for neon and oxygen, can be combined is this manner. The average of these normalized values is 0.95~$\pm$~0.78\%. It is possible to try iteratively different values of U$_e$ in order to match this 0.95 \% effect. One obtains that the difference in U$_e$ between the two phases which would induce this change is 110(90) eV for $^{19}$Ne and 400(320) eV for $^{19}$O. These values, between 20 and 900 eV, are close to the values measur
 ed in different kinds of materials with nuclear reactions \cite{raio}.
\par
This small effect in superconductors is also confirmed in an experiment where $^{74}$As was implanted in tantalum at different temperatures \cite{farkas2009measurement}. For some of the temperatures, although not reported, tantalum was in the superconducting phase. From the reported half-life measured values, a normalized value of 0.22 $\pm$ 0.11\% is obtained. These two values, more than one sigma above zero, can be interpreted as an indication for increased electron screening energy in superconductors of the order of 100 eV.
\par
In conclusion, a high-precision experiment was performed in order to measure half-life and branching ratios of $^{19}$O($\beta^-$) and $^{19}$Ne($\beta^+$) implanted in a niobium foil in its superconducting and metallic phases. No difference was observed between the two phases within the limits of experimental accuracy: 0.04\% for the half-life, 0.5\% and 7\% for the 197/1554 and 197/4377 relative branching ratios of $^{19}$O. The screening potential energy for both nuclei in a superconductor is well below the predicted value of $\sim$20 keV by the superscreening model. An indication for a screening potential energy close of $\sim$100 eV was obtained after combining all measured results. This value is in the same range of values measured by nuclear reactions in different kinds of materials. It is important to measure accurately the screening potential energy in beta decay, in particular in order to determine the systematic errors associated with the high precision measurements made for fundamental physics. 
Recently, the half-life of $^{19}$Ne was used to evaluate the contribution of right-handed currents to the weak interaction \cite{triambak2012high}. In the present study, the simultaneous measurement of half-life and branching ratios limited the beam intensity used. An improved experimental setup dedicated to half-life measurements, based on scintillators with increased efficiency and fast scalers, can be used with a more intense purified radioactive beam. In principle, a factor 10 in accuracy is readily accessible.
\par
We thank the GANIL crew for delivering the radioactives beams, for help with the cryogenic system and friendly collaboration. We are grateful to Dr. A. Navin, Dr. E. Balanzat and Dr. B. Gervais for the interesting discussions.  We acknowledge the support of the European Commission within the Sixth Framework Programme through I3-EURONS (contract no. RII3-CT-2004-506065), the support from the French-Romanian collaboration agreement IN2P3-IFIN-HH Bucharest n$^{\circ}$03-33, the R\'egion Normandie, the support of the French-Serbian CNRS/MSCI collaboration agreement (No 20505) and support of Ministry of Education and Science of Serbia, project No 171018.

\end{document}